\documentclass[draft]{agujournal2019}
\usepackage{url} %this package should fix any errors with URLs in refs.
\usepackage{lineno}
\usepackage[inline]{trackchanges} %for better track changes. finalnew option will compile document with changes incorporated.
\usepackage{soul}
\usepackage{amsmath,amssymb}
\draftfalse

\journalname{JGR-Space Physics}

\begin{document}

\title{Ion Dynamics Across a Low Mach Number Bow Shock}

\authors{D. B. Graham \affil{1}, Yu. V. Khotyaintsev \affil{1}, 
A. P. Dimmock \affil{1}, A. Lalti \affil{1,2}, J. J. Bold\'{u} \affil{1,2}, S. F. Tigik \affil{1}, S. A. Fuselier \affil{3,4} }

\affiliation{1}{Swedish Institute of Space Physics, Uppsala, Sweden.}
\affiliation{2}{Space and Plasma Physics, Department of Physics and Astronomy, Uppsala University, Uppsala, Sweden.}
\affiliation{3}{Southwest Research Institute, San Antonio, TX, USA.}
\affiliation{3}{University of Texas at San Antonio, San Antonio, TX, USA.}

\correspondingauthor{D. B. Graham}{dgraham@irfu.se}

\begin{keypoints}
\item Protons, singly charged helium ions, and alpha particles are observed upstream and downstream of a bow shock crossing. 
\item The alpha particles and helium ions partly obscure the fine structure of the downstream proton distributions. 
\item High time-resolution mass-resolving ion detectors are needed to study the ion dynamics across collisionless shocks. 
\end{keypoints}

%\date{\today}
\begin{abstract}
A thorough understanding of collisionless shocks requires knowledge of how different ion species are accelerated across the shock. We investigate a bow shock crossing using the Magnetospheric Multiscale spacecraft after a coronal mass ejection crossed Earth, which led to solar wind consisting of protons, alpha particles, and singly charge helium ions. The low Mach number of the bow shock enabled the ions to be distinguished upstream and sometimes downstream of the shock. Some of the protons are specularly reflected and produce quasi-periodic fine structures in the velocity distribution functions downstream of the shock. Heavier ions are shown to transit the shock without reflection. However, the gyromotion of the heavier ions partially obscures the fine structure of proton distributions. Additionally, the calculated proton moments are unreliable when the different ion species are not distinguished by the particle detector. The need to high time-resolution mass-resolving ion detectors when investigating collisionless shocks is discussed. 

\end{abstract}

\section*{Plain Language Summary}
One of the ongoing challenges when investigating collisionless shocks is determining the energy partition between electromagnetic fields and different particle species. Resolving this question requires detailed observations of the electromagnetic fields and particle distributions, and is challenging when multiple ion species are present. We investigate a crossing of Earth's bow shock for unusual solar wind conditions; three ion species are observed in the solar wind and behind the bow shock, namely protons, alpha particles, and singly charged helium ions. We investigate the ion dynamics and show that a small fraction of protons are reflected by the electric field associated with the shock, which results in complex ion distributions. However, since the highest time-resolution ion detectors cannot distinguish between different ion species, the heavier ions partly obscure the fine structure of the protons. The heavier ions lead to errors when calculating the bulk properties (e.g., moments) of protons. These observations illustrate the need for high time-resolution ion detectors, which can distinguish different ion species when studying shocks. 

%\begin{article}

\section{Introduction}
Shocks are ubiquitous throughout the universe and are regions of intense particle acceleration. Earth's bow shock is the most studied collisionless shock using spacecraft observations. At the bow shock, the supersonic solar wind is heated and compressed to form the magnetosheath plasma, which convects around Earth's magnetic field. Near Earth, the solar wind is primarily composed of electrons, protons, and alpha particles. Protons and alpha particles crossing the bow shock typically undergo different dynamics \cite{fuselier1994}. In rare cases, significant contributions of singly charged helium ions have been observed following interplanetary shocks \cite{gosling1980,schwenn1980,zwickl1982}. Understanding the dynamics of the different ion species is crucial to understanding how energy is distributed among the different particle species.  

At Earth's bow shock, the plasma is essentially collisionless so the dissipation required to maintain the shock must come from other mechanisms, such as wave-particle interactions \cite{wu1984,wilson2014}. For supercritical shocks, additional dissipation from ion reflection becomes crucial \cite{paschmann1980}. Previous observations have shown that ion reflection is extremely common at the bow shock \cite{paschmann1982} and can occur for subcritical shocks. For quasi-perpendicular shocks, ions reflected by the shock are turned by the magnetic field and accelerated by the convection electric field, so they gain energy and are transmitted through the shock \cite{paschmann1982}. Generally, protons are reflected, while the heavier alpha particles cross the shock without reflection, although there are cases where reflected alpha particles are observed \cite{broll2017}. 

Investigating the fine structure of ion distributions across and downstream of the bow shock has proved difficult from spacecraft observations. Most ion detectors have relied on the spacecraft spin to construct three-dimensional (3D) ion distributions, limiting the cadence to the spin period. Reflected ions have typically been observed in the shock foot \cite{sckopke1983}, while downstream of the shock the reflected ions are observed as an energetic component of the distribution \cite{sckopke1990}. However, with the Magnetospheric Multiscale (MMS) mission, ion distributions are resolved at substantially higher cadence, enabling distributions to be resolved over kinetic scales by the Fast Plasma Investigation (FPI) instrument \cite{pollock2016}. However, FPI does not distinguish between different ion species, making it challenging to investigate the behavior of different ion species. At present, ion detectors that distinguish different ion species operate at low cadences and cannot resolve ion kinetic scales at the bow shock. 

Here we investigate a low Mach number bow shock observed by MMS. Protons, alpha particles, and singly charged Helium ions are observed upstream and downstream of the bow shock. We investigate the dynamics of the three ion species across the bow shock using observations and numerical modelling. We show that the ion dynamics can be investigated due to the low degree of heating across the shock. We discuss the need for high-resolution mass-resolving particle instruments when investigating collisionless shocks. 

\section{Observations} \label{observ}
We use data from the MMS spacecraft; magnetic field ${\bf B}$ data from Fluxgate Magnetometer (FGM) \cite{russell2016}, electric field ${\bf E}$ data from Electric field Double Probes (EDP) \cite{lindqvist2016,ergun2016}, particle distributions and moments from Fast Plasma Investigation FPI \cite{pollock2016}, and ion data from Hot Plasma Composition Analyzer (HPCA) \cite{young2016}. We use both fast and burst mode data. In burst mode FPI measures ion distributions and moments every $150$~ms. 
%The ion detectors of FPI (FPI-DIS) do not distinguish between the different ions components. 
Particles are sorted by energy per charge. In contrast, HPCA measures distributions and moments of different ion species (protons, Helium ions, alpha particles, and Oxygen) but the temporal resolution is limited to $10$~s in both fast and burst mode. Both ion detectors cover all directions when constructing 3D distributions, and as a result lack the angular and energy resolutions to accurately resolve the fast and cold ion distributions of the solar wind. For FPI the angular channels are separated by $11.25^{\circ}$, while the energy resolution is $\Delta E/E \approx 0.13$ with logarithmically spaced energy channels. We primarily use data from MMS1. 

\begin{figure*}[htbp!]
\begin{center}
\includegraphics[width=140mm, height=150mm]{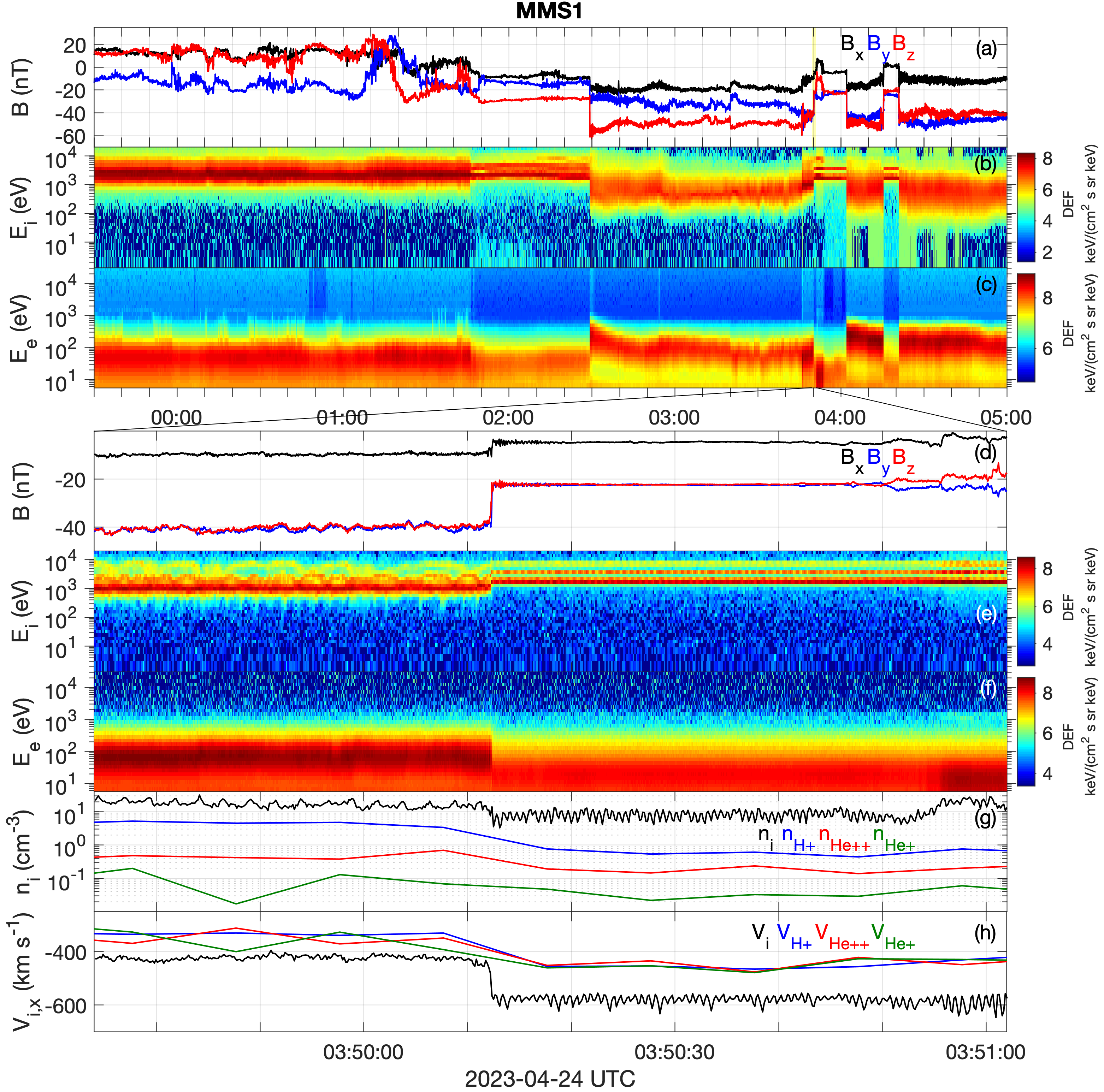}
\caption{Overview of the solar wind and bow shock crossings observed on 24 April 2023 by MMS1. 
(a)--(c) Solar wind and bow shock crossings. (a) ${\bf B}$ in GSE coordinates. (b) Ion omni-directional energy flux. (c) Electron omni-directional energy flux. (d)--(h) Bow shock crossing observed at the time 
indicated by the yellow-shaded region in panel (a). 
(d) ${\bf B}$ in GSE coordinates. (e) Ion omni-directional energy flux. (f) Electron omni-directional energy flux. 
(g) Ion number densities $n_i$ and (h) $x$-component of bulk ion velocities ${\bf V}_i$. The black lines are FPI moments and the blue, red, and green lines are HPCA moments for H$^{+}$, He$^{2+}$, and He$^{+}$, respectively. }
\label{Figure0}
\end{center}
\end{figure*}

%\subsection{Overview}
We investigate a bow shock crossing observed on 24 April 2023 at 03:50:12~UT. The spacecraft were located at $(13.7, -15.9, -9.3)$~$R_E$ in Geocentric Solar Ecliptic (GSE) coordinates, where $R_E$ is Earth's radius. Prior to the bow shock crossings encountered on this day, MMS observed a Coronal Mass Ejection (CME) shock in the solar wind at 17:34:32~UT on 23 April 2023. At $\sim$01:46:20~UT on 24 April 2023 MMS transitioned from the CME sheath to the magnetic cloud, shown in Figures \ref{Figure0}a--\ref{Figure0}c, which provide an overview of the bow shock crossings. The solar wind plasma becomes substantially colder and ${\bf B}$ becomes steady and large amplitude. In the CME sheath and magnetic cloud ${\bf B}$ is unusually large compared with the typical solar wind at 1~AU. After the transition to the magnetic cloud, we observe five bow shock crossings, characterized by sharp changes in ${\bf B}$, broadening of the ion energy fluxes, and increased energy of the electron energy fluxes. During the solar wind magnetic cloud intervals we often see three distinct lines in the ion energy flux (Figure~\ref{Figure0}b), which are most clearly seen just after 03:50~UT. These three peaks correspond to solar wind protons (H$^{+}$) (lowest energy line), alpha particles (He$^{2+}$), and singly charge Helium (He$^+$) ions (highest energy line). Although He$^+$ are generally not observed in the solar wind, He$^+$ can occur in magnetic clouds behind CMEs \cite{schwenn1980,zwickl1982}. 

%We now investigate the second bowshock crossing. Figures \ref{Figure0}d--\ref{Figure0}h provide an overview of this bowshock crossing for the magnetosheath to the solar wind. 
The shock crossing is characterized by a sharp decrease in $|{\bf B}|$, corresponding to the shock ramp. The ion differential energy flux shows three bands upstream of the shock corresponding to H$^{+}$, He$^{2+}$, and He$^{+}$ (Figure \ref{Figure0}e). Downstream of shock the fluxes of He$^{2+}$ and He$^{+}$ exhibit quasi-periodic fluctuations in energy due to their gyromotion. The helium ions have approximately half the frequency of the alpha particle fluctuations due to the lower gyroperiod. The He$^{2+}$ fluxes often overlap in energy with the H$^+$ fluxes. The three flux components remain narrow in energy across the shock, meaning the ion components remain relatively cold downstream of the shock. The electron fluxes show that electrons are heated across the shock (Figure \ref{Figure0}f). 

Figures \ref{Figure0}g and \ref{Figure0}h show the ion density $n_i$ and sunward component of the ion bulk velocity ($V_{i,x}$) measured by FPI and HPCA. The $n_i$ calculated from FPI is significantly larger than the H$^+$ density $n_{H+}$ measured by HPCA. There are significant number densities of He$^{2+}$ and He$^+$ upstream and downstream of the shock. Both FPI and HPCA cannot fully resolve solar wind distributions, so there is significant uncertainty in the density moments. Figure \ref{Figure0}h shows that the ion velocities of H$^+$, He$^{2+}$, and He$^+$ agree with each other across the shock, confirming that there are He$^+$ ions in the solar wind. We note that $V_{i,x}$ calculated from FPI is significantly larger than that calculated from HPCA, primarily due to the He$^{2+}$ and He$^+$ fluxes at higher energies being included in the calculation of the moments, which results in an overestimate of ${\bf V}_i$. 
%These uncertainties in ion moments make it difficult to calculate the shock properties, such as the Alfv\'en Mach number $M_A$ and shock normal angle $\theta_{Bn}$. 

To estimate the H$^+$ velocity we compute the one-dimensional (1D) reduced distributions from FPI and take the velocity where the phase-space density $f_i$ peaks. In the spacecraft frame we obtain ${\bf V}_u = (-510, 40, -100)$~km~s$^{-1}$ upstream of the shock and ${\bf V}_d = (-370, 60, -150)$~km~s$^{-1}$ downstream of the shock in GSE coordinates. These velocities are smaller than the FPI moments, but slightly larger than those from HPCA. Using these velocities and the average upstream and downstream ${\bf B}$ we can define a shock-aligned coordinate system. Using the mixed-mode methods based on changes in ${\bf B}$ and ${\bf V}$ we calculate the shock normal direction to be $\hat{\bf n} = (0.93, 0.10, -0.36)$ in GSE coordinates \cite{abraham-shrauner1972}. From $\hat{\bf n}$ we obtain a shock normal angle of $\theta_{Bn} = 87^{\circ}$, corresponding to a nearly perpendicular shock. Based on four-spacecraft timing of ${\bf B}$ we estimate the shock speed to be $V_{sh} \approx -130$~km~s$^{-1}$ along $\hat{\bf n}$. In the shock frame $V_{u,n} = -320$~km~s$^{-1}$ and $V_{d,n} = -170$~km~s$^{-1}$. 

\begin{figure*}[htbp!]
\begin{center}
\includegraphics[width=140mm, height=150mm]{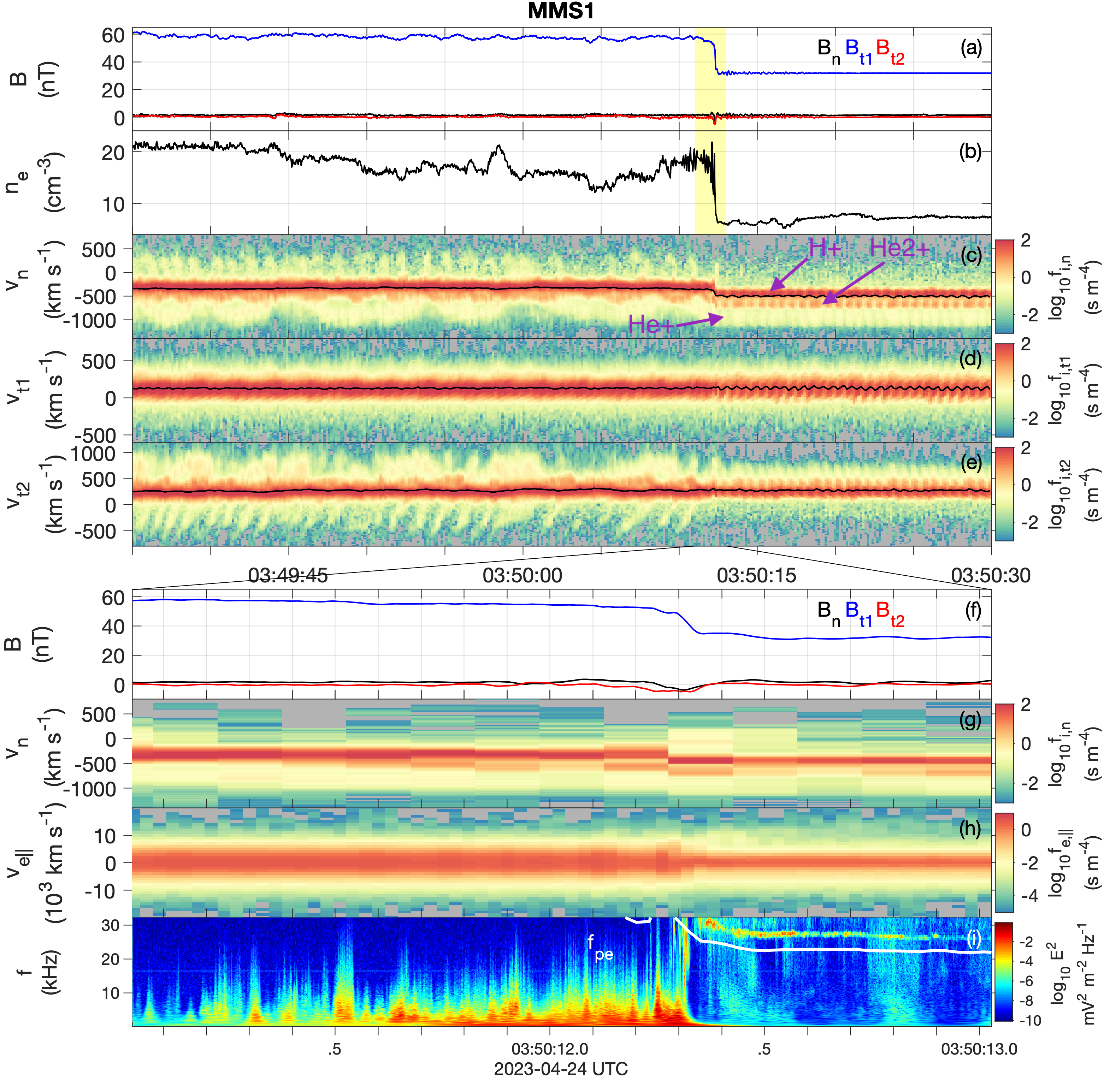}
\caption{Overview of the bow shock crossing observed by MMS1. (a) ${\bf B}$ in ($\hat{\bf n}$,$\hat{\bf t}_1$,$\hat{\bf t}_2$) coordinates. 
(b) $n_e$. (c)--(e) 1D reduced ion distributions in the $\hat{\bf n}$, $\hat{\bf t}_1$, and $\hat{\bf t}_2$ directions, respectively. 
The black lines show the ion bulk velocity calculated from FPI. Panels (f)--(i) show a zoomed in interval across the ramp indicated by the yellow shading in panels (a) and (b). 
(f) ${\bf B}$ in ($\hat{\bf n}$,$\hat{\bf t}_1$,$\hat{\bf t}_2$) coordinates. (g) Reduced ion distributions along $\hat{\bf n}$. (h) 1D reduced electron distributions in the direction parallel to ${\bf B}$. 
(i) Frequency-time spectrogram of ${\bf E}$. The white line is $f_{pe}$ calculated from electron moments.}
\label{Figure1}
\end{center}
\end{figure*}

Figure \ref{Figure1} provides an overview of the bow shock crossing observed by MMS1 in $(\hat{\bf n},\hat{\bf t}_1,\hat{\bf t}_2)$ coordinates, where $\hat{\bf t}_1$ is along the upstream ${\bf B}$, and $\hat{\bf t}_2$ completes the right-hand coordinate system. 
%There is little rotation in ${\bf B}$ across the shock, with $B_n$ and $B_{t2}$ remaining close to zero across the shock (Figure \ref{Figure1}a). 
Figure \ref{Figure1}b shows that there is a sharp increase in the electron number density $n_e$ across the shock coinciding with the increase in $|{\bf B}|$. Figures \ref{Figure1}c--\ref{Figure1}e show the 1D reduced ion distributions $f_i$ along $v_n$, $v_{t1}$, and $v_{t2}$, respectively. To calculate the reduced distributions we convert energy $E$ to speed using $v = \sqrt{2 e E/m_p}$, where $e$ is the unit charge and $m_p$ is the proton mass. This will result in the speeds of He$^{2+}$ and He$^+$ being overestimated by a factor of $\sqrt{2}$ and $2$, respectively. Across the shock we observe that $f_i$ sharply drops to lower $v_n$. In the downstream region we observe the quasi-periodic fluctuations in He$^{2+}$ and He$^+$ in the $v_n$ and $v_{t2}$ components, although both He$^{2+}$ and He$^+$ are observed for $v_n < 0$ and $v_{t2} > 0$. There is significant overlap between the H$^+$ and He$^{2+}$ species. Likewise, the H$^+$ and He$^{+}$ components overlap, although there are intervals where the ions are well separated in velocity space.  Figure \ref{Figure1}d shows that $f_i(v_{t1})$ does not change significantly across the shock. Thus ion heating is largely perpendicular to ${\bf B}$. Additionally, quasi-periodic fine structures in $f_i$ are observed for $v_n > 0$ and $v_{t2} < 0$ downstream of the shock. In Figures \ref{Figure1}c--\ref{Figure1}e we overplot ${\bf V}_i$ computed from FPI. Due to the presence of He$^{2+}$ and He$^+$, ${\bf V}_i$ overestimates the bulk H$^{+}$ speed. This is most evident in Figure \ref{Figure1}c, where $V_{i,n}$ is offset from the peak in $f_i$. 

Figures \ref{Figure1}f--\ref{Figure1}i show the properties of the shock ramp. There is an enhancement in $f_i$ for $v_n > 0$ ahead of the ramp observed over two ion distributions $\sim 0.3$~s (Figure \ref{Figure1}g) due to ion reflection at the ramp. Figure \ref{Figure1}h shows the 1D reduced electron distribution as a function of $v_{e,||}$, where $v_{e,||}$ is the electron speed parallel to ${\bf B}$. Across the shock $f_{e}$ broadens, corresponding to electron heating. Ahead of the ramp we observe weak electron beams at $v_{e,||} \sim 10^4$~km~s$^{-1}$. These are reflected and accelerated electrons, which are typical of quasi-perpendicular shocks. Figure \ref{Figure1}i shows the frequency-time spectrogram of the high-frequency ${\bf E}$ across the shock. The downstream spectrum is characterized by broadband fluctuations with frequency $f$ below $10$~kHz. Just upstream of the shock we observe Langmuir waves at the local electron plasma frequency $f_{pe}$, likely generated by the reflected electrons. We estimate the upstream density from the frequency of the Langmuir waves to be $n_e = 9$~cm$^{-3}$, corresponding to $f_{pe} = 27$~kHz. Thus, $n_e$ is underestimated by FPI upstream of the shock. Assuming flux conservation across the shock, we estimate $n_e = 17$~cm$^{-3}$ downstream of the shock, in agreement with the observed $n_e$. Using these $n_e$ and the scalar electron pressure $P_e$ calculated from FPI, we calculate the upstream and downstream electron temperatures to be $T_e = 10$~eV and $T_e = 37$~eV, respectively. We estimate the Alfv\'en and fast magnetosonic Mach numbers to be $M_A \approx 1.4$ and $M_f \approx 1.4$, which are very low values for Earth's bow shock \cite{lalti2022}. Due to the presence of multiple ion species and the unreliability of the temperature measurements for very cold ion populations, it is difficult to calculate meaningful ion temperatures $T_i$ for this event, although $T_i$ does not significantly affect the calculation of $M_f$ in this case. The shock properties correspond to a subcritical shock \cite{kennel1985}, so ion reflection is expected to be minimal. 

\section{Ion motion}
We investigate the motion of the different ion species across the bow shock using a test-particle model. We perform the calculations in the normal incidence frame (NIF), and transform the particle motion to the spacecraft frame for direct comparison with observations. To model the bow shock we assume a perpendicular shock with ${\bf B}$, $n_e$, and $P_e$ profiles given by
\begin{linenomath}
\begin{equation}
B_{t1}(n) = - B_0 \tanh{\left( \frac{n}{l} \right)} + B_1, 
\label{Bmodel}
\end{equation}
\begin{equation}
n_e(n) = -n_0 \tanh{\left( \frac{n}{l} \right)} + n_1, 
\label{nmodel}
\end{equation}
\begin{equation}
P_e(n) = -P_0 \tanh{\left( \frac{n}{l} \right)} +P_1, 
\label{Pemodel}
\end{equation}
\end{linenomath}
where $l$ is the characteristic width of the ramp, $B_0$, $B_1$, $n_0$, $n_1$, $P_0$, and $P_1$ determine the jump conditions, and $n$ is the distance along $\hat{\bf n}$. To model the observed shock we use $B_0 = 13.5$~nT, $B_1 = 45.5$~nT, $n_0 = 4$~cm$^{-3}$, $n_1 = 13$~cm$^{-3}$, $P_0 = 4.25 \times 10^{-2}$~nPa, $P_1 = 5.75 \times 10^{-2}$~nPa, and $l = 10$~km. The current density ${\bf J}$ is then given by
\begin{linenomath}
\begin{equation}
J_{t2}(n) = - \frac{B_0 \mathrm{sech}^2 \left( \frac{n}{l}  \right)}{\mu_0 l}.
\label{Jmodel}
\end{equation}
\end{linenomath}
The electric field along $\hat{\bf n}$ and $\hat{\bf t}_2$ is given by
\begin{linenomath}
\begin{equation}
E_n = - \frac{1}{e n_e(n)}\left( J_{t2}(n) B_{t1}(n) + \frac{\partial P_e(n)}{\partial n} \right), 
\label{Enmodel}
\end{equation}
\begin{equation}
E_{t2} = - \frac{V_{u,n}(n_1 - n_0) B_{t1}(n)}{n_e(n)},
\label{Et2model}
\end{equation}
\end{linenomath}
where $V_{u,n}$ is the upstream speed in the NIF. The gradient in $P_e$ contributes to only $\approx 10$~\% of the total $E_n$. Equation (\ref{Et2model}) determines the average ${\bf V}_i$ upstream and downstream of the shock via ${\bf E} \times {\bf B}$ drift. The particle velocities and trajectories are calculated by integrating the usual equations of motion
\begin{linenomath}
\begin{equation}
\frac{d {\bf x}}{d t} = {\bf v}, \frac{d {\bf v}}{d t} = \frac{Z e}{m_i} \left( {\bf E} + {\bf v} \times {\bf B} \right),
\label{equationmotion}
\end{equation}
\end{linenomath}
where $Z$ is the charge number of the ion and $m_i$ is the ion mass. 

\begin{figure*}[htbp!]
\begin{center}
\includegraphics[width=140mm, height=150mm]{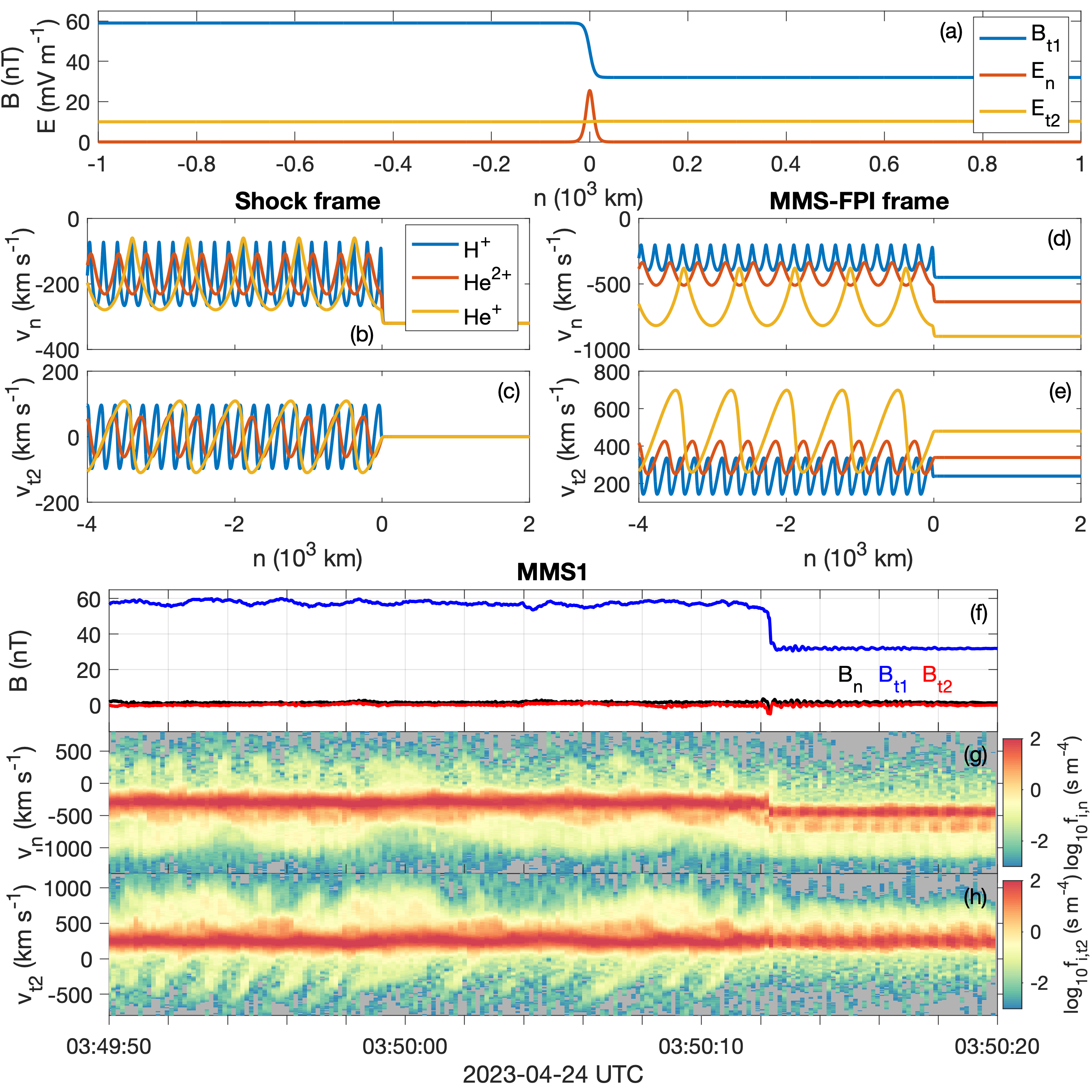}
\caption{Ion motion across the model bow shock. (a) Profiles of $B_{t1}$, $E_n$, and $E_{t2}$. 
(b) $v_n$ versus $n$ and (c) $v_{t2}$ versus $n$. 
(d) $v_n$ versus $n$ and (e) $v_{t2}$ versus $n$ in the 
MMS-FPI frame. The blue, red, and yellow lines indicate H$^+$, He$^{2+}$, and He$^+$, respectively. 
(f) ${\bf B}$ observed by MMS1. (g) and (h) 1D reduced ion distributions along $v_n$ and $v_{t2}$.}
\label{Figure2}
\end{center}
\end{figure*}

Figure \ref{Figure2}a shows the profiles of the model ${\bf B}$ and ${\bf E}$. For these shock conditions $E_{t2}$ is uniform across the shock. In Figures \ref{Figure2}b and \ref{Figure2}c we plot the velocities $v_n$ and $v_{t2}$ of H$^+$, He$^{2+}$, and He$^{+}$ versus $n$ in the NIF. For each species the upstream initial condition is $v_n = V_{u,n} = -320$~km~s$^{-1}$, so there is no force on the particles upstream of the shock and ${\bf v}$ is constant. At the ramp there is sudden drop in $v_n$ due to the cross-shock potential $\phi = - \int E_n dn = 512$~V. After this decrease in $v_n$ due to $\phi$ the ions begin cyclotron turning because they are no longer moving at the ${\bf E} \times {\bf B}$ velocity. This initially results in a further decrease in $v_n$ and $v_{t2} < 0$. After this point the ions undergo periodic gyromotion, with $v_n$ remaining negative while $v_{t2}$ fluctuates around $0$. 
%This drop in velocity can be approximated using 
%\begin{equation}
%\frac{1}{2} m_p \left( V_{sw,n}^2 - V_{d,n}^2 \right) = e Z \phi.
%\label{dvphi}
%\end{equation}
%From equation (\ref{dvphi}) we estimate the decreases in speed across the ramp to be $250$~km~s$^{-1}$, $90$~km~s$^{-1}$, and $40$~km~s$^{-1}$ for H$^+$, He$^{2+}$, and He$^{+}$, respectively. 

We now consider how the ion motion will be observed in the spacecraft frame using FPI. This conversion is given by
\begin{linenomath}
\begin{equation}
{\bf v}_{i,sc} = \sqrt{\frac{Z m_i}{m_p}} \left( {\bf v}_i + {\bf V}_{sh} \right),
\label{vframe}
\end{equation}
\end{linenomath}
where ${\bf V}_{sh}$ = (-130, 120, 240)~km~s$^{-1}$ is the velocity of the NIF in ($\hat{\bf n}$,$\hat{\bf t}_1$,$\hat{\bf t}_2$) coordinates relative to the spacecraft. The dependence on $m_i$ and $Z$ result from assuming the H$^+$ mass and charge when converting particle energy to speed. The results of equation (\ref{vframe}) are shown in Figures \ref{Figure2}d and \ref{Figure2}e. The frame transformation shifts the ions apart in velocity compared with the NIF where all the ions overlap. Upstream of the shock, the ions are well separated in velocity allowing them to be distinguished by FPI. Downstream of the shock there are intervals where the different ions are well separated from each other in velocity space. In particular, He$^+$ is largely separated from the other ions. Figures \ref{Figure2}g and \ref{Figure2}h show the reduced distributions along $v_n$ and $v_{t2}$ for comparison. Qualitatively, these periodic motions in He$^{2+}$ and He$^+$ are seen in the reduced distributions, along with the separation between the ions upstream of the shock.

%This drop in velocity can be approximated by $V_{i,n} = \sqrt{2 Z e \phi/m_p}$, which corresponds to $310$~km~s$^{-1}$, $220$~km~s$^{-1}$, and $160$~km~s$^{-1}$ for H$^+$, He$^{2+}$, and He$^+$, respectively. 

%For the bulk of the incoming ions to be reflected by the shock potential $\phi$ the ions must satisfy $1/2 m_i V_{i,n}^2 < Z e \phi$. From the model we calculate a cross-shock potential of $\phi = 510$~V, while the proton kinetic energy in the NIF is $\approx 535$~eV. Thus, the shock potential can only reflect a small fraction of protons. 

\section{Ion distributions} 
We now extend the test-particle model to investigate the predicted distributions downstream of the shock. To model the ion distributions downstream of the shock we use Liouville's theorem, which states that the phase-space density is constant along particle trajectories. This is expressed as 
\begin{linenomath}
\begin{equation}
f({\bf x}_1,{\bf v}_1) = f({\bf x}_0,{\bf v}_0), 
\label{liouvilleeq}
\end{equation}
\end{linenomath}
where ${\bf x}_1$ and ${\bf v}_1$ are solutions to equation (\ref{equationmotion}), and ${\bf x}_0$ and ${\bf v}_0$ refer the initial conditions, the solar wind in this case. We assume that the fields governing the equations of motion and upstream solar wind distributions do not change in time, so we can neglect the time dependence. We assume the solar wind ion distributions are Maxwellian. In this model, there are no forces acting on the ions along $\hat{\bf t}_1$ so we use a two-dimensional (2D) reduced distribution for the solar wind given by
\begin{linenomath}
\begin{equation}
f_i(v_n,v_{t2}) = \frac{n_i}{\pi v_i^2} \exp{\left( - \frac{(v_n - V_{sw,n})^2 + v_{t2}^2}{v_i^2} \right)}, 
\label{swdist}
\end{equation}
\end{linenomath}
where $v_i = \sqrt{2 k_B T_i/m_i}$ is the ion thermal speed and $k_B$ is Boltzmann's constant. Figure \ref{Figure1}d shows that there is little change in the distribution along $v_{t1}$ across the bow shock, supporting the use of the reduced 2D distribution for the model. For the solar wind we assume a H$^{+}$ temperature of $T_p = 3$~eV and that He$^{2+}$ and He$^{+}$ have the same thermal speed as H$^{+}$ ($T_i = 12$~eV). For the H$^{+}$ number density we use $n_{H+} = 8$~cm$^{-3}$ and for He$^{2+}$ and He$^+$ we assume densities of $n_{He2+} = 0.1 n_{H+}$ and $n_{He+} = 0.01 n_{H+}$. We compute $f({\bf x}_1,{\bf v}_1)$ by integrating backwards in time using ${\bf x}_1$ and ${\bf v}_1$ as initial conditions. For trajectories reaching the upstream, equations (\ref{liouvilleeq}) and (\ref{swdist}) are used to compute $f({\bf x}_1,{\bf v}_1)$, otherwise $f({\bf x}_1,{\bf v}_1) = 0$ is used. 

\begin{figure*}[htbp!]
\begin{center}
\includegraphics[width=160mm, height=120mm]{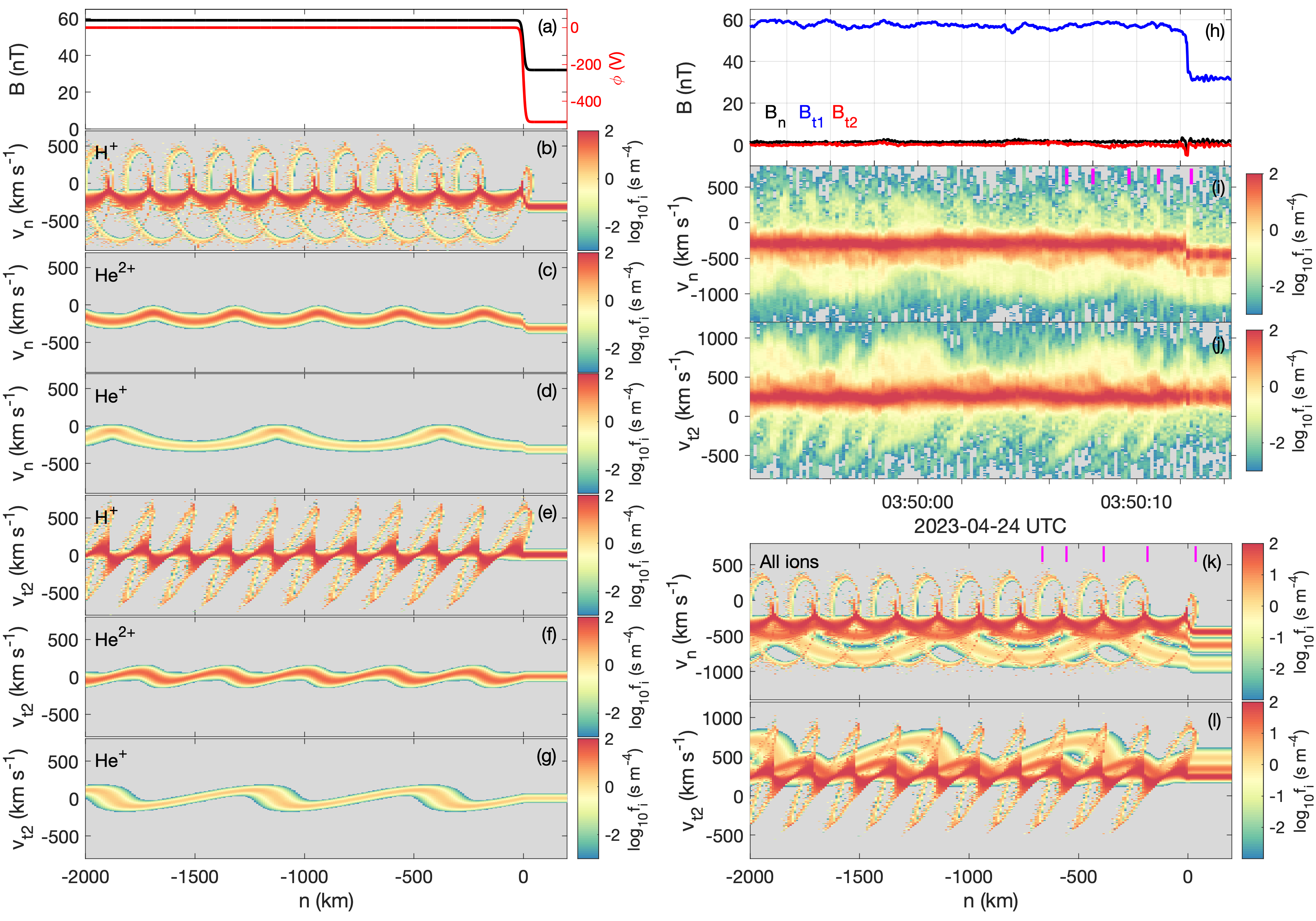}
\caption{Ion distributions across the shock predicted from Liouville mapping and comparison with distributions observed by MMS1. (a) Profile of $B_{t1}$ and $\phi$ of the model bow shock. (b)--(d) 1D reduced distributions along $v_n$ for H$^+$, He$^{2+}$, and He$^+$, respectively. (e)--(g) 1D reduced distributions along $v_{t2}$ for H$^+$, He$^{2+}$, and He$^+$, respectively. These distributions are plotted in the NIF. (h)--(j) ${\bf B}$, 1D reduced distributions along $v_n$ and $v_{t2}$ observed by MMS1. (k) and (l) Modelled 1D reduced distributions along $v_n$ and $v_{t2}$ in the MMS-FPI frame. The H$^+$, He$^{2+}$, and He$^+$ distributions are converted to energies measured by FPI.}
\label{Figure3}
\end{center}
\end{figure*}

The resulting distributions of H$^{+}$, He$^{2+}$, and He$^+$ across the shock are shown in Figure \ref{Figure3}. Figures \ref{Figure3}b--\ref{Figure3}d show the  distributions along $v_n$ in the NIF for H$^+$, He$^{2+}$, and He$^+$, and Figures  \ref{Figure3}e--\ref{Figure3}g show the distributions along $v_{t2}$ in the same format. The H$^{+}$ distributions downstream of the shock consist of directly transmitted H$^{+}$ and a small fraction of reflected H$^{+}$ ($\sim 10$~\% in the model). The H$^+$ transmitted without reflection undergo periodic fluctuations along $v_n$ and $v_{t2}$. The reflected H$^+$ are specularly reflected by $\phi$ into the upstream region where they are accelerated by the convection electric field $E_{t2}$, and have significantly higher velocities when they return to the shock and are transmitted across the ramp. These reflected H$^{+}$ are responsible for the fine structures seen in Figures \ref{Figure3}b and \ref{Figure3}e, and are consistent with those observed by FPI (Figures \ref{Figure3}i and \ref{Figure3}j). We note that there are some differences between the observed reflected H$^+$ and in the model. In observations for $v_n > 0$ we see hook-shaped structures rather than the full loop associated with reflected H$^+$. For $v_{t2} < 0$ we observe two narrow bands associated with reflected H$^+$ that meet at the largest values of $|v_{t2}|$ in the model, while in observations we only observe a single line. These differences may result from the angular and energy resolution of FPI, which may not fully resolve these fine structures. 

For He$^{2+}$ and He$^+$ the distributions are decelerated by $\phi$ and compressed across the ramp. The shock potential is too small to reflect these ions. The distributions gyrate downstream of the shock, which is consistent with observations. Additionally, the densities and temperatures fluctuate and peak where $v_n$ is closest to $0$, primarily due to  broadening in $f_i$ along $v_{t2}$. This results in non-gyrotropic ion distributions. The He$^{2+}$ and He$^+$ distributions averaged over multiple gyrations result in ring-like distributions in the $v_n$--$v_{t2}$ plane. 

In Figures \ref{Figure3}k and \ref{Figure3}l we convert the ion distributions to the MMS-FPI frame and overlay the three species together to model the observed distributions. When compared with the observed distributions by FPI over a comparable spatial scale (Figures \ref{Figure3}i and \ref{Figure3}j), we find very good agreement between observations and the model. For He$^{2+}$ and He$^+$ the transformation broadens the ion distributions, and is most evident for He$^+$. Figure \ref{Figure3}k shows that along $v_n$ the He$^+$ distribution overlaps with the reflected H$^+$ for $v_n < 0$, obscuring the observation of the fine structures associated with reflected H$^+$. Similarly, along $v_{t2}$ the He$^+$ distribution overlaps with the reflected H$^+$ for $v_{t2} > 0$. The He$^{2+}$ distributions obscure the periodic fluctuations in the distribution of transmitted H$^+$. 

\begin{figure*}[htbp!]
\begin{center}
\includegraphics[width=160mm, height=75mm]{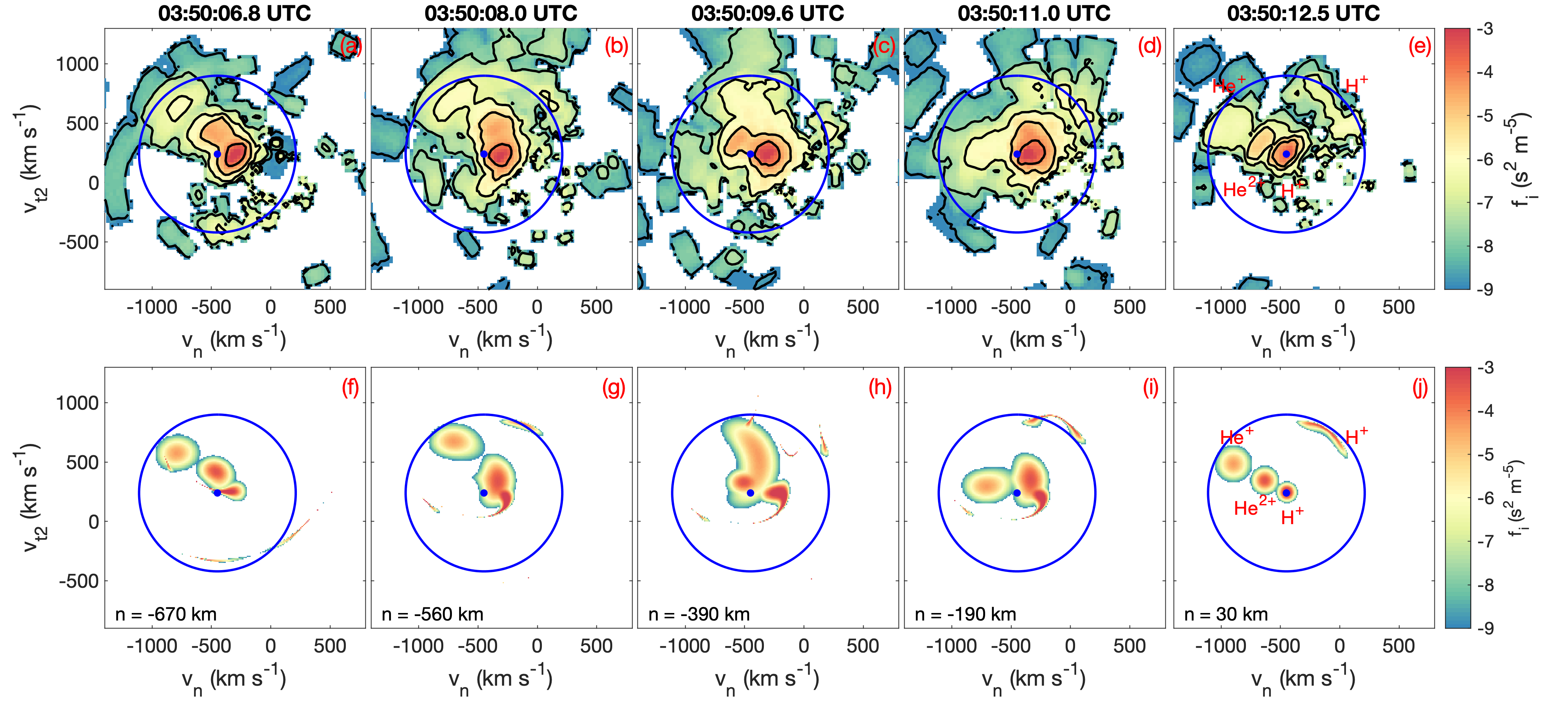}
\caption{2D ion distributions in the $v_{n}$--$v_{t2}$ plane in the MMS spacecraft reference frame. (a)--(e) Ion distributions observed by FPI at the times indicated by the magenta lines in Figure \ref{Figure3}i. (f)--(j) Model ion distributions in the MMS-FPI frame at the locations indicated by the magenta lines in Figure \ref{Figure3}k. Model distributions present all ion species. (e) and (j) correspond to the shock foot, while the remaining distributions are behind the ramp with increasing distances from right to left. The blue dots indicate the solar wind speed and the blue circle indicates the velocities where specularly reflected H$^+$ should be observed. }
\label{Figure4}
\end{center}
\end{figure*}

To further investigate the ion distributions downstream of the shock we compare observed 2D reduced distributions in the $v_n$--$v_{t2}$ plane with those predicted from the model, shown in Figure \ref{Figure4}. Figures \ref{Figure4}a--\ref{Figure4}e show ion distributions observed by FPI at the times indicated by the magenta lines in Figure \ref{Figure3}i, while Figures \ref{Figure4}f--\ref{Figure4}j show model distributions at the positions indicated by the magenta lines in Figure \ref{Figure3}k. For direct comparison, the model distributions are presented in the MMS-FPI frame and taken from positions corresponding to the observed distributions. Figures \ref{Figure4}e and \ref{Figure4}j show the ion distributions in the foot, where reflected H$^+$ are observed. In observations and the model the reflected H$^+$ occur in the same region of velocity space and have velocities in good agreement with the theoretical prediction of speeds of $|{\bf v} - {\bf V}_{u}| = 2 V_{u,n}$ for specularly reflected H$^+$ in the NIF (indicated by the blue circles in Figure \ref{Figure4}). The incoming ion distributions approximately correspond to the solar wind ion distributions, and are relatively well separated from each other. 

Figures \ref{Figure4}a--\ref{Figure4}d and \ref{Figure4}f--\ref{Figure4}i show distributions just downstream of the shock. In each case the relative positions of the H$^+$, He$^{2+}$, and He$^{+}$ distributions agree between observations and the model. Figures \ref{Figure4}d and \ref{Figure4}i show distributions around where the He$^{2+}$ temperature is expected to peak. The H$^+$ and He$^{2+}$ distributions overlap significantly, while the He$^{+}$ distribution is relatively well separated from the other distributions. Figures \ref{Figure4}c and \ref{Figure4}h show distributions where the He$^+$ temperature is expected to peak. The model shows a crescent-shaped distribution of He$^+$, which overlaps with the H$^+$ and He$^{2+}$ distributions. The observed distribution shows the same overlaps with the H$^+$ and He$^{2+}$ distributions, although the crescent shape of He$^+$ is not resolved. The remaining distributions show significant overlap between H$^+$ and He$^{2+}$, while the He$^+$ distribution is more separated from these distributions. 

In each of the observed distributions we see the reflected H$^+$ in different regions of velocity space. Comparing with the model we see that these reflected H$^+$ occur around the same velocities as predicted by the model. In the model some of the reflected H$^+$ occur in or near the same region of velocity space as the He$^+$ distributions, specifically in Figures \ref{Figure4}f, \ref{Figure4}h, and \ref{Figure4}i. Thus, the He$^{+}$ ions partly obscure some of the reflected H$^+$ in the observed distributions. 

In summary, we find excellent agreement between the observed and modeled ion distributions. A small fraction of H$^+$ are specularly reflected at the shock due to $\phi$, resulting in quasi-periodic fine structures in the observed ion distributions. In contrast, He$^{2+}$ and He$^{+}$ transit the shock without any reflection. Quasi-periodic fluctuations in He$^{2+}$ and He$^{+}$ are observed, which are due to the gyromotion of these ions downstream of the shock. These gyrating ions are observed by MMS and partly obscure the reflected H$^+$. 

\section{Discussion}
We now discuss the effects of He$^{2+}$ and He$^+$ on the particle moments downstream of the shock. When computing ion moments with FPI it is generally assumed that H$^+$ are the dominant or only ion species and thus heavier ions do not significantly affect the calculation. The effect of the heavier ions on ${\bf V}_i$ was evident in Figure \ref{Figure0}, where FPI overestimated the bulk velocity. In particular, the sunward ion speed estimated from the ion moments in the solar wind is $V_{sw,x} = -580$~km~s$^{-1}$, while we estimate ${\bf V}_{sw,x} = -510$~km~s$^{-1}$ from the peak in $f_i$. This overestimate of ${\bf V}_{sw}$ calculated from the particle moments will result in overestimated Mach numbers and could influence estimates of $\theta_{Bn}$ using mixed-mode methods \cite{abraham-shrauner1972}. However, reasonable estimates of ${\bf V}_{sw}$ can be obtained by isolating the energy ranges associated with H$^+$, when the solar wind is relatively cold. Similarly, ${\bf V}_i$ is overestimated by the moments downstream of the shock. 

Additionally, ion temperatures downstream of the shock become extremely difficult to estimate due the significant overlap of the different ions. Figure \ref{Figure4} shows that the H$^{+}$, He$^{2+}$, He$^{+}$ distributions typically overlap, and this overlap is most significant at the positions where the He$^{2+}$ and He$^{+}$ temperatures peak according to the model. Since the contributions from He$^{2+}$ and He$^{+}$ are significant for this shock the temperature computed from FPI will likely not match the actual H$^{+}$. We can consider this effect by comparing with the model predictions. From FPI we obtain a median scalar ion temperature of $T_i \approx 50$~eV (for simplicity we discuss the scalar temperature rather than the full temperature tensor). From the model the median $T_i$ of all H$^+$ is $\approx 120$~eV. This high $T_i$ is due to the reflected H$^+$ and thus $T_i$ depends strongly on the fraction of reflected H$^+$. From the model, reflected H$^+$ constitute $\approx 16$~\% of the total average downstream $n_{H+}$. The fraction of reflected H$^+$ is very sensitive to $\phi$ and the properties of the upstream H$^{+}$ distribution, such as bulk velocity, temperature, and the shape of the distribution function. If we only consider the H$^+$ transmitted through the shock without reflection we obtain $T_i \approx 8$~eV, compared with upstream $T_i = 3$~eV. Thus, there is only a small amount of heating associated with transmitted H$^+$. 

For He$^{2+}$, $T_i$ ranges from $16$~eV to $29$~eV, with a median value of $18$~eV, while for He$^{+}$, $T_i$ ranges from $13$~eV to $59$~eV, with a median value of $17$~eV. For both He$^{2+}$ and He$^{+}$, $T_i = 12$~eV is assumed for upstream distributions. The peaks in $T_i$ are due to gyrophase bunching and occur where the ion distributions overlap with each other in observations (Figure \ref{Figure4}), so it is difficult to verify these $T_i$ changes in observations. Finally, we can calculate the model $T_i$ for the combined ion distributions (Figures \ref{Figure3}k and \ref{Figure3}l) in the spacecraft frame. For these distributions the median $T_i$ is $120$~eV, while if we neglect reflected H$^+$ we obtain $T_i \approx 35$~eV. Thus, the case without reflected H$^+$ is a closer, albeit lower, approximation to the observed $T_i$. This might suggest that in the model the fraction of reflected H$^+$ could be overestimated compared with observations. However, it is also possible that FPI does not fully resolve the reflected H$^+$, which occur in narrow regions of velocity space, because of the finite angular and energy resolution of the instrument. From the above discussion we conclude that He$^{2+}$ and He$^{+}$ contribute to the measured $T_i$ (typically assumed to be H$^+$) and illustrates the need for high-resolution mass resolving particle instruments, as was proposed for ESA's THOR mission \cite{vaivads2016} and for Plasma Observatory \cite{retino2022}. 
 
This bow shock crossing is an unusual case because three ion species are resolved and $M_A$ is very low. Thus $T_i$ remains relatively small allowing the different ion species to be partly resolved downstream of the shock by FPI. For Earth's bow shock typical Mach numbers are $5 \lesssim M_A \lesssim 10$ \cite{lalti2022}, so ion heating will be substantially stronger. Moreover, these shocks are typically highly dynamic, so the quasi-periodic fluctuations in the ion distributions are unlikely to occur. Thus, different ion species would not be resolved at ion kinetic scales. 

Finally, around the ramp we observe intense electrostatic waves reaching $\sim 300$~mV~m$^{-1}$. These waves are consistent with ion-acoustic-like waves, which can potentially interact with ions. The fact that the fine structures in the ion distributions associated with the reflected H$^+$ remain well downstream of the ramp suggests that the electrostatic waves do not play a substantial role in ion scattering. The ion dynamics are well explained by the laminar ${\bf B}$ and ${\bf E}$ of the shock. 

\section{Conclusions}
In this paper we have investigated the ion dynamics of a low Mach number bow shock crossing, where protons, alpha particles, and singly charged helium ions are observed upstream and downstream of the shock. 
The key results are: 
\begin{itemize}
    \item Highly structured proton distributions are observed downstream of the bow shock and are caused by a small fraction of the incoming protons being reflected by the cross-shock potential. The gyromotion of the reflected protons results in quasi-periodic striations in the 1D reduced ion distributions. 
    \item Alpha particles and singly charged helium ions do not undergo reflection but are compressed and heated across the bow shock. 
    \item The gyromotion of the alpha particles and singly charged helium ions downstream of the shock partly obscures the fine structures associated with the proton distributions. 
\end{itemize}

These results demonstrate the need for high time-resolution mass resolving particle instruments when investigating collisionless shocks in space. At present, mass resolving particle detectors onboard spacecraft cannot resolve ion distributions associated with shocks at or below ion kinetic scales. Thus, the complicated ion distributions associated with reflected protons cannot be resolved. At present the highest time-resolution ion detector in space, namely FPI onboard MMS, does not distinguish between different ion species. Thus, when significant concentrations of heavier ions are present, such as He$^{2+}$ and He$^{+}$, these ions prevent accurate proton moments from being calculated. 

\section*{Data Availability Statement}
MMS data are available at https://lasp.colorado.edu/mms/sdc/public.

\acknowledgments
We thank the entire MMS team and instrument PIs for data access and support. 
This work is supported by the European Union's Horizon 2020 research and innovation program under grant agreement number 101004131 (SHARP). 

%\end{article}

%\bibliography{magrecpapers}

\end{document}